\documentclass[11pt,twoside]{article}

\usepackage{asp2006}
\usepackage{epsf}
\usepackage{psfig}
\usepackage{lscape}

\begin{document}
\title{Galactic Disk Formation and the Angular Momentum Problem}   
\author{Andreas Burkert}   
\affil{University Observatory Munich, Scheinerstr. 1, D-81679 Munich, Germany}    

\begin{abstract} 
Galactic disk formation requires
knowledge about the initial conditions under which disk galaxies form,
the boundary conditions that affect their secular evolution and
the micro-physical processes that drive the multi-phase interstellar medium
and regulate their star formation history. Most of these ingredients are still
poorly understood. Recent high-resolution observations of young
high-redshift disk galaxies provide insight into early phases of galactic disk formation
and evolution. Combined with low-redshift disk data these observations should 
eventually allow us to reconstruct the origin and evolution of
late-type galaxies. I 
will summarize some of the major problems that need to 
be addressed for a more consistent picture of
galactic disk formation and evolution.
\end{abstract}


\section{Initial- and Boundary Conditions: The Cosmological Angular Momentum Problem}

Galaxy formation to some extent is an initial condition problem.
Whether a galactic disks can form at all depends on the amount
of angular momentum present in the infalling gas. The disk structure
is determined by the gravitational potential of the baryonic and dark component of the
galaxy and the specific angular momentum
distribution of the fraction of infalling gas that can cool and dissipate its potential 
and kinetic energy while settling into centrifugal equilibrium in the equatorial plane.
Once a massive disk has formed and on timescales longer than the 
infall timescale
secular disk evolution will become important, resulting in
angular momentum redistribution of gas and stars in the disk by viscous effects
and gravitational torques, coupled with star formation and selective
gas loss in galactic winds (Kormendy \& Kennicutt 2004). 

The origin of angular momentum is generally believed to be cosmological.
Before and during the early phase of protogalactic collapse,
gas and dark matter are well mixed
and therefore acquire a similar specific angular momentum distribution
(Peebles 1969; Fall \& Efstathiou 1980; White 1984). 
If angular momentum would be conserved during gas
infall, the resulting disk size should be directly related to the 
specific angular momentum $\lambda'$ of the surrounding dark halo where
(Bullock et al. 2001)

\begin{equation}
\lambda' = \frac{J}{\sqrt{2} M_{vir} V_{vir} R_{vir}}
\end{equation}

\noindent with $R_{vir}$ and $V_{vir}^2=GM_{vir}/R_{vir}$ the virial radius and 
virial velocity of the halo, respectively, and $M_{vir}$ its virial mass.
Adopting a flat rotation curve, the disk scale length is
(Mo et al. 1998; Burkert \& D'Onghia 04)

\begin{equation}
R_d \approx 8 \left( \frac{\lambda'}{0.035} \right) \left(\frac{H_0}{H} \right)
\left( \frac{v_{max}}{200 km/s} \right) kpc
\end{equation}

\noindent where $v_{max}$ is the maximum rotational velocity in the disk
and $H$ is the Hubble parameter that depends on cosmological redshift z as

\begin{equation}
H = H_0 \left[ \Omega_{\Lambda} + \Omega_M(1+z)^3 \right]^{1/2} .
\end{equation}

\noindent Typical values for a standard $\Lambda$CDM cosmology are
$H_0$ = 73 km/s/Mpc, $\Omega_M$ = 0.238 and $\Omega_{\Lambda} = 0.762$.

\begin{figure}
\includegraphics{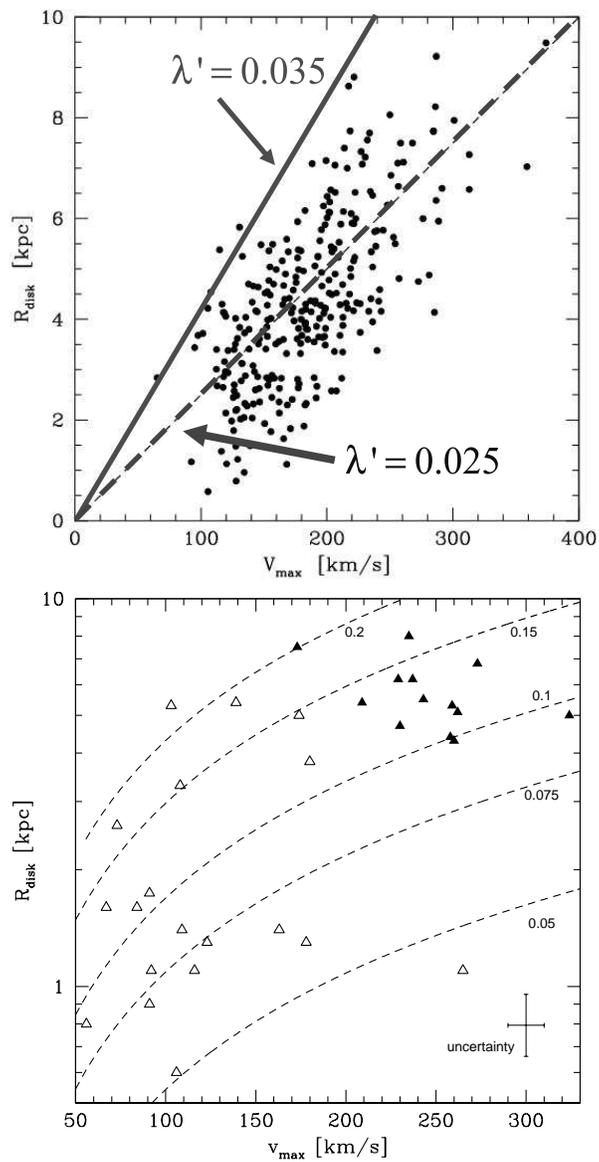}
\caption{The upper panel shows the observed scale lengths versus the maximum rotational velocities of
galactic disks for the Courteau (1997) sample. The solid line shows the 
theoretically predicted correlation for $\lambda' = 0.035$. 
The dashed curve corresponds to $\lambda' = 0.025$. The lower panel shows the disk scale
length versus the maximum velocity of the SINS high-redshift disk sample.
Open and filled triangles correspond to dispersion-dominated and rotation-dominated
galaxies, respectively. Note that here the vertical axis is plotted logarithmically
for better resolution of the dispersion-dominated galaxies.
Dashed curves show the Mo, Mao \& White models with spin parameters
as indicated by the labels.}
\end{figure}

The upper panel of figure 1 shows the correlation between the disk scale length
$R_{disk}$ and the maximum rotational velocity $v_{max}$ for massive spiral galaxies (Courteau 1997) 
There is a linear relationship
which can be fitted well by $\lambda' = 0.025$ which is somewhat smaller than the theoretically 
predicted value of $\lambda' = 0.035$,  indicating that the gas 
could on average have lost some amount of angular momentum during the infall phase.

This result is promising. The situation is however more confusing if we look at the recent
high-resolution observations of z=2 star forming disk
galaxies (F\"orster-Schreiber et al. 2006, 2009; Genzel et al. 2006, 2008; Cresci et al. 2009,
Burkert et al. 2009). In contrast to their
low-redshift counterparts, the high-z galaxies are characterized
by high gas velocity dispersions of $\sigma \approx 40 - 80 km/s$.
In addition, the
sample segregates strongly into two distinct classes at a critical value
of $v_{max}/\sigma \approx 3$. One can
empirically define dispersion-dominated galaxies
as objects with $v_{max}/\sigma \leq 3$ while rotation-dominated galaxies 
are defined by $v_{max}/\sigma > 3$.  The lower panel of figure 1 
shows the half-light radii $r_{1/2}$ of the SINS high-redshift galaxies
versus their maximum rotational velocity $v_{max}$. 
Most of the dispersion-dominated galaxies (open triangles in figure 1) have radii of order 1-2 kpc 
and rotational velocities of order 100 km/s while the radii and rotational velocities
of the rotation dominated galaxies (filled triangles)
are on average a factor of 2-3 larger. The correlation between scale length and velocity of
this sample is however very similar to the low-redshift galaxies.
According to equation (1), this  requires
a factor 3-4 larger spin parameter (dashed curves in the lower
panel of figure 1) for high-redshift disks as a result of the fact that at z=2 the dark halo
virial radii are a factor 3-4 smaller. Burkert et al. (2009) argue that
this could be explained by turbulent pressure effects in the disk and
a less centrally concentrated dark halo component that did not experience adiabatic contraction
during disk formation.

Simulations of galactic disk formation suffer often from 
catastrophic angular momentum loss which 
leads to disks with unreasonably small scale lengths and surface densities that are too large.
The origin might be strong clumping of the infalling gas 
which looses angular momentum by dynamical friction within the surrounding dark matter halo 
(Navarro \& Benz 1991, Navarro \& Steinmetz 2000), low numerical
resolution (Governato et al. 2004, 2007), substantial and major mergers (d'Onghia et al. 2006) 
and artificial secular angular momentum transfer from the cold disk to its hot surrounding 
(Okamoto et al. 2003). It has been argued that this problem might be solved by
including star formation and energetic feedback (e.g. Sommer-Larsen et al. 2003, Abadi et al. 2003,
Springel \& Hernquist 2003, Robertson et al. 2004, Oppenheimer \& Dave 2006, Dubois \& Teyssier 2008).
No reasonable, universally applicable feedback 
prescription has however yet been found that would lead to the formation of large-sized, 
late-type disks, not only for special cases, but in general.

Recently Zavala et al. (2008) showed
that the specific angular momentum distribution
of the disk forming material follows closely the angular momentum evolution of the dark matter halo.
The dark matter angular momentum grows at early times as a result of large-scale tidal torques, 
consistent with the prediction of linear theory and remains constant after the epoch of maximum 
expansion. During this late phase angular momentum is redistributed within the dark halo with
the inner dark halo regions loosing up to 90\% of their specific angular momentum to
the outer parts which is probably related to minor 
mergers with mass ratios less than 10:1.
It is then likely that any gas residing in the inner regions during such
an angular momentum redistribution
will also loose most of its angular momentum, independent of whether the gas resides already
in a protodisk, is still confined to dark matter substructures or is in an extended,
diffuse distribution.
Zavala et al. (2008) (see also Okamoto et al., 2005 and Scannapieco et al., 2008) show that
efficient heating of the gas component can prevent angular momentum loss, probably
because most of the gaseous component resides in the outer parts of the dark halo
during its angular momentum redistribution phase. The gas would then actually gain angular
momentum  rather than loose it and could lateron settle smoothly into
an extended galactic disk in an ELS-like (Eggen, Lynden-Bell \& Sandage 1962) accretion phase.

Little is known about the energetic processes that
could lead to such an evolution. Obviously, star formation must be delayed during the
protogalactic collapse phase in order 
for the gas to have enough time to settle into the plane before condensing into stars. 
However star formation is also required in order to heat the gas, preventing
it from collapsing prior to the angular momentum redistribution phase.
Scannapieco et al. (2008)
show that their supernova feedback prescription is able to regulate star formation while at the same time pressurizing
the gas. Their models are however still not efficient enough in order to produce disk-dominated,
late-type galaxies. Large galactic disks are formed. The systems are however dominated by a
central, massive, low-angular momentum stellar bulge component. This is in contradiction with
observations which indicate a large fraction of massive disk galaxies with bulge-to disk
ratios smaller than 50\% (Weinzirl et al. 2008) that cannot be produced currently by numerical simulations
of cosmological disk formation.

\section{Energetic Feedback and Star Formation}

As argued in the last section star formation and energetic feedback plays a dominant
role in understanding the origin and evolution of galactic disks and in determining the
morphological type of disk galaxies. Scannapieco et al. (2008) for example demonstrate that the
same initial conditions could produce either an elliptical or a disk galaxy, depending
on the adopted efficiency of gas heating during the protogalactic collapse phase.
A consistent model of the structure and evolution of the multi-phase, turbulent 
interstellar medium and its condensation into stars is still missing. This situation is however
improving rapidly due to more sophisticated numerical methods and fast computational
platforms that allow us to run high-resolution models, incorporating a large number of possibly
relevant physical processes (Wada \& Norman 2002, Krumholz \& McKee 2005, Tasker \& Bryan 2008, 
Robertson \& Kravtsov 2008).

Cosmological simulations often adopted simplified
observationally motivated descriptions of star formation that are based
on the empirical Kennicutt relations (Kennicutt 1998, 2007) that come in two different version.
The first relation (K1) represents a correlation between the star formation rate per surface area
$\Sigma_{SFR}$ and the gas surface density $\Sigma_g$, averaged over the whole galaxy

\begin{equation}
\Sigma_{SFR}^{(K1)} = 2.5 \times 10^{-4} \left(\frac{\Sigma_g}{M_{\odot}/pc^2}\right)^{1.4} 
\frac{M_{\odot}}{kpc^2 \ yr}
\end{equation}

\noindent The second relation (K2) includes a dependence on the typical orbital period
$\tau_{orb}$ of the disk

\begin{equation}
\Sigma_{SFR}^{(K2)} = 0.017 \left(\frac{\Sigma_g}{M_{\odot}/pc^2}\right)
\left( \frac{10^8 yrs}{\tau_{orb}} \right) \frac{M_{\odot}}{kpc^2 \ yr}.
\end{equation}

\noindent These relationships have been derived from observations as an average over
the whole disk. They are however often also used as theoretical prescriptions for the local
star formation rate which appears observationally justified if 
the total gas surface densitiy $\Sigma_g$ is replaced by the local surface density of molecular gas.
The origin of both relationships is not well understood yet. 
One can combine K1 and K2 and derive a relationship between the average gas density in 
galactic disks and their orbital period

\begin{equation}
\Sigma_g \sim \tau_{orb}^{-2.5} \sim \left( \frac{v_{rot}}{R_{disk}} \right)^{2.5}
\end{equation}

\noindent where $v_{rot}$ and $R_{disk}$ are the rotational velocity and the size of the galactic disk,
respectively.
This result is puzzling as it is not clear why the kinematical properties of galactic disks
should correlate with their gas surface densities especially in galaxies of Milky Way type or earlier 
where the gas fraction is small compared to the mass in stars ( see e.g.
Robertson \& Kravtsov 2008).

\section{Secular Evolution and Turbulence in Galactic Disks}

Dark halos have a universal
angular momentum distribution that should also be characteristic for the infalling gas component
(Bullock et al 2001).
Van den Bosch et al. (2001) lateron showed that this angular momentum distribution is
not consistent with the observed distribution of exponential galactic disks.
One possibility is angular momentum redistribution during filamentary cold gas infall
(Dekel et al. 2009). Another suggestion is viscous disk evolution.
The viscosity is likely driven by interstellar turbulence which is a result of
stellar energetic feedback processes or global disk instabilities (magneto-rotational instability
or gravitational instability).
Viscous effects will however increases the angular momentum problem substantially as viscosity in general
removes angular momentum from the dominante mass component in the disk and transfers it to the outermost
parts of the disk.

Slyz et al. (2002) studied the viscous formation of exponential stellar disks from gas disks with 
various different surface density distributions. Their numerical simulations show that
exponential disks form if the star formation timescale is of order
the viscous timescale. Genzel et al. (2008) derive a timescale for turbulent viscosity in galactic
disks of (see also Dekel et al. 2009)

\begin{equation}
\tau_{visc} = \frac{1}{\alpha} \left( \frac{v_{rot}}{\sigma} \right)^2 \tau_{orb}
\end{equation}

\noindent where $\alpha$ is of order unity. $\tau_{visc} \approx 10^{10}$ yrs for
disks like the Milky Way with $\sigma \approx$ 10-20 km/s and self-regulated low star 
formation rates. $z \sim 2$ star forming disk galaxies on the other hand are 
characterized by large random
gas motions of order 40 km/s to 80 km/s and viscous timescales of less than $10^9$yrs
(Genzel et al., 2006, 2008, F\"orster-Schreiber et al. 2006).
Interestingly, for these objects, the star formation timescales are again similar to the
viscous timescale, leading to star formation rates of 100 M$_\odot$/yr and confirming
that galactic disk gas turbulence, star formation and secular evolution are intimately coupled.

Burkert et al. (2009) find a good correlation between the observed irregular gas
motion in high-redshift disk galaxies and the theoretically expected gas velocity dispersion, adopting
a Toomre Q parameter of unity.  This is expected if the main driver of clumpiness and turbulence
is gravitational disk instability. A disk that is kinematically too cold
with small velocity dispersions is highly gravitationally unstable. Gravitational instabilites
generate density and velocity irregularities that drive turbulence and heat
the system kinematically. The gas
velocity dispersion increases till it approaches the stability limit characterised
by Q=1  where kinetic driving by gravitational
instabilities saturates. A disk with even higher velocity dispersions would be stable,
turbulent energy would
dissipate efficiently and the velocity dispersion would decrease again until it crosses
the critical velocity dispersion limit at Q=1 where gravitational instabilities become efficient again in
driving turbulent motions. More simulations are required to investigate this process
in greater details.

\section{Summary}

Many complex non-linear processes affect
galactic disk formation and evolution. We are still in an early stage
of understanding these processes. One of the most
interesting questions is how the observations of high-redshift disk
galaxies can be combined with galaxies in the local universe. The high-z disks generate stellar
systems with high velocity dispersions that resemble more rotating early-type 
or S0 galaxies than Milky-Way type objects. Where are then the progenitors
of present-day disks? In addition, why did their progenitors evolve differently
with respect to the observed high-z galaxies? How do bulge-less disk galaxies form?
Given the recent progress both in observations and theoretical modelling the
time seems ripe to solve these questions.

\acknowledgements I would like to thank Shardha Jogee for inviting me to this
interesting conference, her great hospitality and for financial support.

\end{document}